\documentclass{ifacconf} 
\usepackage{graphicx}    
\usepackage{natbib}
\usepackage{blindtext}
\usepackage{amsmath}
\usepackage{placeins}
\usepackage{amsfonts}
\usepackage{caption}
\usepackage{amssymb}
\usepackage{mathtools}

\usepackage{color,soul}
\begin{document}

\begin{frontmatter}

\title{Remote State Estimation with Privacy Against Eavesdroppers } 

\author[First]{Matthew J.\ Crimson}
\author[First]{Justin M.\  Kennedy} 
\author[First]{Daniel E.\ Quevedo}

\address[First]{School of Electrical Engineering and Robotics, Queensland University of Technology, Brisbane QLD, 4000 Australia.
(e-mail: matthew.crimson@hdr.qut.edu.au; \{j12.kennedy,daniel.quevedo\}@qut.edu.au)}

\begin{abstract}             
We study the problem of remote state estimation in the presence of a passive eavesdropper, under the challenging network environment of no packet receipt acknowledgments. A remote legitimate user estimates the state of a linear plant from the state information received from a sensor via an insecure and unreliable network.
The transmission from the sensor may be intercepted by the eavesdropper.
To maintain state confidentiality, we propose an encoding scheme.
Our scheme transmits noise based on a pseudo-random indicator, pre-arranged at the legitimate user and sensor. The transmission of noise harms the eavesdropper's performance, more than that of the legitimate user. Using the proposed encoding scheme, we impair the eavesdropper’s expected estimation performance, whilst minimising expected performance degradation at the legitimate user.
We explore the trade-off between state confidentiality and legitimate user performance degradation.

\end{abstract}


\end{frontmatter}

\section{Introduction}
Cyber-Physical Systems (CPS) have proliferated through modern society with the ubiquitous rise of the Internet of Things. CPS integrate cyber infrastructure and physical systems and encompass large-scale, geographically dispersed systems \citep{ishii2022security}. CPS have extensive applications in advanced manufacturing, transportation, and energy. Due to the tight integration of cyber and physical components in such systems, CPS are vulnerable to cyber attacks. Cyber vulnerabilities can be exploited by malicious adversaries, which seek to disrupt CPS operations. Attacks on CPS may result in performance degradation or even system failure \citep{humayed2017cyber}. 

Security of a system is defined by availability, integrity, and confidentiality. Attacks on CPS can therefore be classified into three classes which target these security goals; denial-of-service (DoS), deception attacks and eavesdropping, which target the security goals of availability, integrity, and confidentiality respectively \citep{cardenas2008secure}. The wireless communications of state information and control actions are susceptible to adversaries intercepting measurements through eavesdropping. There is a need to design estimation algorithms over wireless communication networks that account for the wireless network, while also ensuring confidentiality of the transmitted information from an eavesdropper. Guaranteeing confidentiality is crucial for sensor and control data, which convey information about the physical system state.

Remote estimation schemes deal with time-critical systems. Encoding schemes can utilise acknowledgments from a remote user to encode transmissions at the sensor, utilising an acknowledgment channel. These encoding techniques are lightweight and avoid significant computation \citep{tsiamis2018state}. Cryptography-based tools \citep{katz2020introduction} are at times used in practice, however, they introduce computation and communication overheads \citep{lee2010price} and therefore are only of limited use. However, utilising an acknowledgement channel is a point of vulnerability for eavesdropping and DoS attacks \citep{ding2020remote}, as experimentally demonstrated with the jamming of acknowledgments utilising commodity hardware \citep{klingler2019jamming}. Therefore, the motivation for this work is to design a low-complexity encoding scheme in a system without acknowledgments by adapting transmissions to ensure confidentiality of the state.

In this paper, we consider a sensor transmitting the
state estimate of a stable, linear first-order dynamic system to a legitimate user over a random packet dropping wireless network. A malicious adversary may perform an eavesdropping attack and intercept the transmission, compromising state confidentiality. We design an encoding scheme that is activated by the detection of an eavesdropper.
\section{Remote State Estimation with an Eavesdropper} \label{sec:problem formulation}
We consider a dynamic system that is modelled as a first-order scalar, discrete-time invariant system
\begin{equation} \label{eq:lti}
 x_{k+1} = ax_{k} + w_k,  \quad y_{k} = x_{k} + v_k ,
\end{equation}
where $x_k \in \mathbb{R}$, for time $k \geq 0$ with stable dynamics $|a| < 1$, measured by a sensor with $y_k \in \mathbb{R}$ with noises $w_k \in \mathbb{R}$ and $v_k \in \mathbb{R}$. The noises $w_k$ and $v_k$ are independent, uncorrelated, identically distributed (i.i.d) zero-mean, Gaussian noise with variances $q \geq 0$, $r > 0$ respectively, with $\mathbb{E}[w_kv_k] = 0$. The initial state of the process $x_0$ is a Gaussian random variable with zero mean and variance $\mathbb{E}[x_0^2] = \sigma_0$. The initial state is uncorrelated with $w_k$ and $v_k$ with $\mathbb{E}[x_0w_k] = 0$ and $\mathbb{E}[x_0v_k] = 0$. All system and noise parameters $(a, q, r, \sigma_0)$ are assumed to be \textit{public knowledge}, available to the sensor, legitimate user and eavesdropper.

In current CPS frameworks, sensors are equipped with on-board processors to improve the estimation
performance of the system. Sensors process the collected information by executing recursive algorithms, enabled by advanced embedded systems-on-chip \citep{ding2017multi}. To emulate a smart sensor configuration, this paper considers the sensor configuration is computationally capable of optimal state estimation in real-time. The sensor locally estimates the state $x_k$, based upon all measurements it has collected up to time $k$ and transmits its local estimate to the remote user over a wireless network. Denote $\hat{x}_{k}^{s}$ as the sensor's local Minimum Mean-Squared Error (MMSE) estimate of state, $x_{k}$, and $P_k^{s}$ as the corresponding estimation error variance
\begin{equation*}
    \begin{split}
    \hat{x}_k^s &= \mathbb{E}[x_k|y_0, y_1,...,y_k] \\
        {P}_k^s &=\mathbb{E}[(x_k-\hat{x}_k^s)^2|y_0, y_1,...,y_k]. 
    \end{split}
\end{equation*}
A standard Kalman filter can be utilized to provide an optimal MMSE state estimate. For any initial condition, the error variance $P_k^s$ converges exponentially fast to some unique value $\Bar{P}$ \citep{anderson2012optimal}.
Without loss of generality, it is assumed the local state estimator has entered steady state operation and that $P_k^s = \Bar{P}$, where $\bar{P}$ is the unique solution to 
\begin{equation} \label{eq:pbar}
    \bar{P} = a^2\bar{P}+q - (a^2\bar{P}+q)^2(a^2 \bar{P}+q+r)^{-1} .
\end{equation}
The legitimate user requires to reliably estimate the system state for the purposes of remote monitoring or control. To obtain this estimate, the sensor transmits a packet $z_k \in \mathbb{R}$ of state information or noise. We define the formation of $z_k$ in Section \ref{sec:Transmissions at the Sensor}. The packet $z_k$ is transmitted over a wireless network to the legitimate user. However, the transmitted packets can also be received by an eavesdropper. We utilize a standard packet based transmission utilized in network control problems. Denote the packet reception indicator outcome at the legitimate estimator by $\lambda_k \in \{0, 1\}$, where
\begin{equation*}
    \lambda_k = \begin{cases}
        1, \quad \textrm{if a packet is received at the legitimate user} \\
        0, \quad \textrm{if a packet dropout occurs.}
    \end{cases}
\end{equation*}
Denote the packet reception indicator outcome at the eavesdropper by $\lambda_k^e \in \{0, 1\}$, where
\begin{equation*}
    \lambda_k^e = \begin{cases}
        1, \quad \textrm{if a packet is received at the eavesdropper} \\
        0, \quad \textrm{if a packet dropout occurs.}
    \end{cases}
\end{equation*}
The channel outcomes for the legitimate user and eavesdropper are modelled as i.i.d Bernoulli processes, independent of the initial state of the process and the process noise. Let the channel quality be defined as the probability of packet dropout. The probability of packet dropout for the legitimate user's channel is defined as
\begin{equation} \label{eq:legitchannel}
    \bar{\gamma} = \mathbb{P}[\lambda_k = 0]
\end{equation}
Similarly, the probability of packet dropout for the eavesdropper's channel is defined as
\begin{equation} \label{eq:eavesdropperchannel}
    \gamma^e = \mathbb{P}[\lambda_k^e = 0]
\end{equation}
We introduce the concept of ``open loop performance'' for later discussion as upper bounds for the eavesdropper's and legitimate user's estimation error variance. Open loop estimation occurs in the absence of information from the sensor. It is characterised by $ x^{OP}_k = \mathbb{E}[x_k] = 0, \ P^{OP}_k = \mathbb{E}[x_k^2]$, where the asymptotic open loop estimate  and estimation error variance satisfy:
\begin{equation} \label{eq:openloop}
 x^{OP} \triangleq \lim_{k\to\infty}x_k^{OP}= 0 , \quad P^{OP} \triangleq \lim_{k\to\infty}P^{OP}_k=\frac{q}{1-a^2},
\end{equation}
which is the unique solution to the geometric series of the Lyapunov equation for the scalar case with $|a| < 1$ \citep{anderson2012optimal}. 
\section{Transmissions at the Sensor} \label{sec:Transmissions at the Sensor}
\begin{figure}
    \centering
    \includegraphics[width=8.4cm]{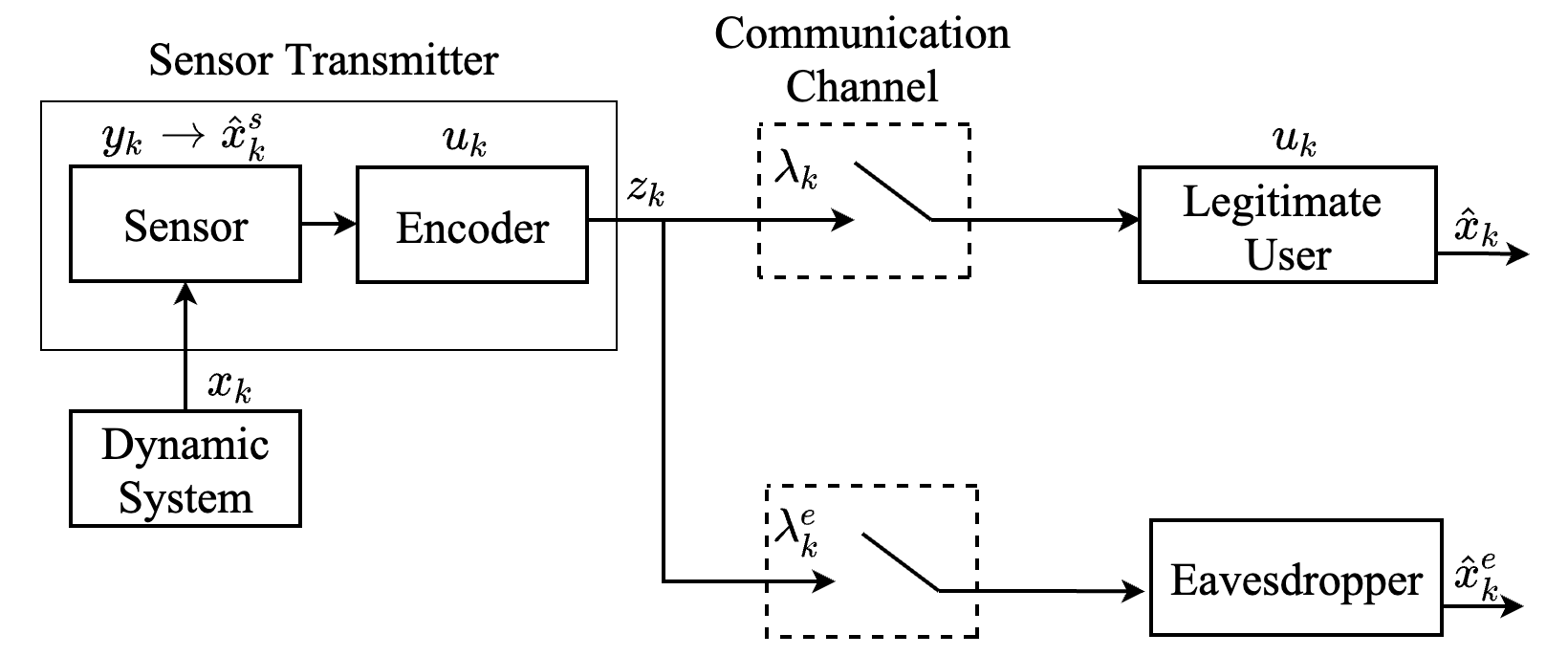}
    \caption{The sensor transmits $z_k$, an encoded packet formed by the indicator $u_k$. The indication to send noise $u_k$ is known to the legitimate user and sensor, and unknown to the eavesdropper. The packet is transmitted over a packet dropping network. The legitimate estimator and eavesdropper compute state estimates $\hat{x}_k$ and $\hat{x}_{k}^e$. }
    \label{fig:block}
\end{figure}
The considered remote estimation architecture is shown in Fig. \ref{fig:block} and consists of a sensor observing a dynamic system, a legitimate user, and an eavesdropper. The sensor transmits packets of information over a wireless channel. To degrade the eavesdropper’s performance, we desire to limit the amount of useful information it receives. To actively harm its performance, we propose to send noise. 

Our proposed strategy is to send either the sensor's state estimate $\hat{x}_k^s$ or noise $n_k$ by using a Bernoulli pseudo-random indicator $u_k \in \{0, 1\}$. The transmission of noise reduces transmissions of information to the legitimate user, reducing its performance. We pre-arrange the indicator $u_k$ between the legitimate estimator and sensor. This encoding scheme forms $z_k$ which is the transmitted packet from the sensor
\begin{equation} \label{eq:scheme}
z_k = 
\begin{cases} 
\hat{x}_k^s, & \text{ if } u_k = 1 \\
n_k, & \text{ if } u_k = 0. \\
\end{cases}
\end{equation}
The indicator $u_k$ informs what is transmitted at the sensor. In practical systems, random numbers are generated through pseudo-random algorithms. By agreeing upon an initial seed at the legitimate user and sensor, we pre-arrange the indicator sequence $u_k$.

We denote the probability of the sensor transmitting noise as 
\begin{equation} \label{eq:muchoice}
    \mu = \mathbb{P}(u_k = 0).
\end{equation}
This probability constitutes the main design variable. It can be adjusted based on the dropout channel probabilities and the known process noise of the dynamics to balance the impact between degrading the legitimate user's performance and ensuring data confidentiality. We explore this trade-off in Section \ref{sec:design issues}.
 
We choose the noise to have the same characteristics as the state $x_k$. The noise is designed such that a change in transmission cannot be easily detected at the eavesdropper, as it exhibits the same characteristics as the state $x_k$ \citep{naha2022quickest}. The noise $n_k$ is normally distributed, with zero mean $\mathbb{E}[n_k] = 0$ and variance $\mathbb{E}[n_k^2] = q$ uncorrelated with the state $x_k$ where $\mathbb{E}[x_kn_k] = 0$. 

Under this formulation, the goal is to decide how often to send noise such that data confidentiality is protected. The indicator is designed to be pseudo-random, such that the eavesdropper cannot easily discern patterns in transmission. Moreover, we require the sensor's state estimate to be received by the legitimate user intermittently to ensure its performance is not degraded by successive losses of state information. Due to the pre-arrangement of the indicator $u_k$, the need for an acknowledgment channel is omitted. 

We denote $\mathcal{I}_k$ as a collection of historical information at the legitimate user up to time $k$, with $\mathcal{I}_k \triangleq \{u_0,\lambda_0,\lambda_0z_0,...,u_k,\lambda_k,\lambda_kz_k\}, \mathcal{I}_{-1}\in \O$. We define the legitimate user's own state estimate and the corresponding estimation error variance as
\begin{equation} \label{eq:legituser}
    \begin{split}
    \hat{x}_k &\triangleq \mathbb{E}[x_k|\mathcal{I}_k] \\ P_k &\triangleq \mathbb{E}[(x_k-\hat{x}_k)^2|\mathcal{I}_k],
    \end{split}
\end{equation}
which is based on all sensor data packets received up to time step $k$. We denote $\mathcal{I}^e_k$ as a collection of historical information at the eavesdropper up to time $k$ with $\mathcal{I}_k^e \triangleq \{\lambda_0^e, \lambda_0^ez_0,...,\lambda_k^e,\lambda_k^ez_k\}, \mathcal{I}_{-1}^e\in \O$. We define the eavesdropper's own state estimate and the corresponding estimation error variance as
\begin{equation} \label{eq:eaves}
    \begin{split}
    \hat{x}^e_k &\triangleq \mathbb{E}[x_k|\mathcal{I}_k^e]\\
    P_k^e &\triangleq \mathbb{E}[(x_k-\hat{x}_k^e)^2|\mathcal{I}_k^e],
    \end{split}
\end{equation}
which is based on all sensor data packets received up to time step $k$. The eavesdropper is unaware of the indication for the sensor to send noise $u_k$ and can incorporate noise $n_k$ into its state estimate.
\section{Eavesdropper Mitigation Strategy} \label{sec:Remote State Estimation and Performance}
In an  unreliable, insecure network, the state estimate may not be received by the legitimate user, or the received packet may be noise due to the employed encoding scheme. To estimate the state of the system with intermittent observations, a second estimation scheme is employed at the legitimate user.

The legitimate user obtains the state estimate as follows: Once the sensor’s transmission arrives, and the sensor has not transmitted noise, the estimator synchronizes $\hat{x}_k$ with that of the sensor. Otherwise, the legitimate user predicts $x_k$ based on the previous estimate using the system model (\ref{eq:lti}). Thus, the state estimate at the legitimate user $\hat{x}_k$ \eqref{eq:legituser} satisfies
\begin{equation*}
    \hat{x}_k = \begin{cases}
        \hat{x}_k^s, \quad ~~~~~ \text{ if } (\lambda_k,u_k) = (1,1)\\
        a\hat{x}_{k-1}, ~~~ \text{ if } \lambda_k = 0 \text{ or }  (\lambda_k,u_k) = (1,0).
    \end{cases} 
\end{equation*}
The corresponding estimation error variance $P_k$ satisfies
\begin{equation} \label{eq:estlegitimate}
    P_k = \begin{cases}
        \bar{P}, \quad ~~~~~~~~~~ \text{ if }  (\lambda_k,u_k) = (1,1)\\
        a^2P_{k-1} + q, ~ \text{ if } \lambda_k = 0 \text{ or } (\lambda_k,u_k) = (1,0).
    \end{cases} 
\end{equation}
It follows from \eqref{eq:estlegitimate} that when a packet arrives and it is the state estimate from the sensor, the legitimate user's estimation error variance is that of the sensor's $\bar{P} $ \eqref{eq:pbar}. Under successive packet dropouts or transmissions of noise from the sensor as $k \rightarrow \infty$, the estimation error variance of the legitimate user has an upper bound of open loop prediction of $P^{OP}$ \eqref{eq:openloop}.

The goal of the eavesdropper is to construct an estimate of the state of the system $x_k$  based on all sensor data packets received up to time step $k$. The eavesdropper is unaware of the  randomised indicator to send noise (see \eqref{eq:scheme}). Hence, the eavesdropper cannot distinguish $\hat{x}_k^s$ from $n_k$. This feature can be used to deceive the eavesdropper to incorporate noise $n_k$ into its state estimate.  The eavesdropper obtains the state estimate as follows: Once the sensor’s transmission arrives, and the sensor has not transmitted noise, the eavesdropper synchronizes $\hat{x}^e_k$ with that of the sensor. If the sensor’s transmission arrives, and the sensor has transmitted noise, the eavesdropper synchronizes $\hat{x}^e_k$ with the transmitted noise $n_k$.  If a packet dropout occurs, the eavesdropper predicts $x_k$ based on the previous estimate using the system model \eqref{eq:lti}, such that the eavesdropper's state estimate $\hat{x}_k^e$ \eqref{eq:eaves} from the sensor's viewpoint satisfies
\begin{equation} \label{eq:eveasdropperestimation}
    \hat{x}_k^e = \begin{cases}
        \hat{x}_k^s, \quad ~~ \text{ if } (\lambda_k^e,u_k) = (1,1)\\
        n_k, \quad ~~ \text{ if } (\lambda_k^e,u_k) = (1,0)\\
        a\hat{x}_{k-1}^e, \text{ if } \lambda_k^e = 0. 
    \end{cases} 
\end{equation}
We characterise the eavesdropper's estimation error variance in Lemma \ref{lemma:Estimation error variance}.
\begin{lem} \label{lemma:Estimation error variance}
The  estimation error variance $P_k^e$ \eqref{eq:eaves} satisfies
\begin{equation} \label{eq:esteavesdropper}
    P_k^e = \begin{cases}
        \bar{P}, \quad ~~~~~~~~~~\text{ if } (\lambda_k^e,u_k) = (1,1)\\
        P_n, \quad ~~~~~~~~~ \text{ if } (\lambda_k^e,u_k) = (1,0) \\
        a^2P_{k-1}^e + q, ~ \text{ if } \lambda_k^e = 0,
    \end{cases}
\end{equation}
where $P_n$ is 

\begin{equation} \label{eq:covuppereavesdropper}
    P_n = P^{OP} + q,
\end{equation}

which is the estimation error variance under receiving and utilizing a packet of noise, as the packet is mistaken for the state estimate, with $P^{OP}$ defined in \eqref{eq:openloop}.

\begin{pf}
The proof is included in the Appendix.
\end{pf}
\end{lem}
Lemma \ref{lemma:Estimation error variance} characterises the instantaneous worst-case performance for the eavesdropper under the encoding scheme. It shows that if noise is transmitted often, we can drive the eavesdropper's estimation error variance above open loop prediction. However, greater transmissions of noise degrade the legitimate user's performance as well. Therefore, the probability of $\mu$ in \eqref{eq:muchoice} needs to be chosen with care. We explore this trade-off in the following section.

\section{Design Issues} \label{sec:design issues}
To quantify the estimation performance of the legitimate user and eavesdropper under the encoding scheme from a sensor viewpoint, we explore the mathematical expectation of the estimation error variance. The use of the expectation allows us to compute the expected performance over the stochastic channel and encoding scheme. Particularly, due to the absence of acknowledgments, the channel outcomes are unknown at the sensor. Thus we use the expected estimation performance to design the sensor's encoding variable $\mu$, see \eqref{eq:muchoice}. We assume that the legitimate user's channel quality 
$\bar{\gamma}$ and the eavesdropper's channel quality $\gamma^e$ are both known.

We present an analytic expression of the expectation of estimation error variance of the legitimate user in Lemma \ref{lemma:estimation error legitimate}  and of the eavesdropper in Lemma \ref{lemma:estimation error eavesdropper}. Our expressions depend on  the channel qualities \eqref{eq:legitchannel}, \eqref{eq:eavesdropperchannel}, process dynamics~\eqref{eq:lti}, and the encoding design variable \eqref{eq:muchoice}.
\begin{lem} 
\label{lemma:estimation error legitimate}
The expectation of the legitimate user's estimation error variance satisfies
\begin{equation} \label{eq:esterrorlegitimatefull}
    \lim_{k\to\infty}\mathbb{E}[P_k] = \frac{\bar{P}(1-\bar{\gamma})(1-\mu) +q(\bar{\gamma} + (1-\bar{\gamma})\mu)}{1 - a^2(\bar{\gamma} + (1-\bar{\gamma})\mu)} .
\end{equation}
\end{lem}
\begin{pf}
The proof is included in the Appendix.
\end{pf}
\begin{lem} \label{lemma:estimation error eavesdropper}
The expectation of the eavesdropper's estimation error variance obeys
\begin{equation} \label{eq:esterroreavesdropperfull}
    \lim_{k\to\infty}\mathbb{E}[P_k^e] = \frac{\bar{P}(1-\gamma^e)(1-\mu) + q\gamma^e + P_n(1-\gamma^e)\mu}{1 - a^2\gamma^e} .
\end{equation}
\end{lem}
\begin{pf}
The proof is included in the Appendix.
\end{pf}
Inspecting the result of Lemma 3, we derive the following result on the expected performance of the eavesdropper as a function of the encoding scheme.
\begin{lem} \label{lemma:mu}
    Consider any eavesdropper channel quality $0\leq \gamma^e<1$, dynamics \eqref{eq:lti}, and suppose that the design variable $\mu$ in \eqref{eq:muchoice} is chosen as
    \begin{equation}
            \mu^{OP} = \frac{P^{OP}(a^2\gamma^e -1) -\gamma^e\bar{P} + \gamma^eq + \bar{P}}{(\gamma^e -1)(P_n - \bar{P})}.
    \label{eq:mudesignforopenloop}
    \end{equation}
    We then have:
    $$\lim_{k\to\infty}\mathbb{E}[P_k^e] = P^{OP},$$ where $P^{OP}$ is defined in \eqref{eq:openloop}.
\end{lem}
\begin{pf}
The proof is included in the Appendix.
\end{pf}
Considering the choice of  $\mu$  in our encoding scheme, we desire to meet two conditions. Firstly, the legitimate user's expected estimation error variance should be upper bounded by that of  open-loop prediction
\begin{equation*}
     \lim_{k\to\infty}\mathbb{E}[P_k] < P^{OP}.
\end{equation*}
Secondly, the eavesdropper’s expected estimation error variance should be greater than open loop prediction
\begin{equation*}
    \lim_{k\to\infty}\mathbb{E}[P_k^e] > P^{OP}.
\end{equation*}
These conditions ensure data confidentiality against an eavesdropper by driving the eavesdropper's estimation error variance above open loop prediction. This means the expected eavesdropper's estimation performance when using received packets is worse than if it had not used any packets. Moreover, confidentiality comes at a cost to the legitimate user's performance, and we desire some of the transmissions to provide sufficient useful information. The following theorem   provides a range for $\mu$ that protects data confidentiality, whilst keeping transmissions informative to the legitimate user.
\begin{thm} \label{thm:secrecy}
 Consider any legitimate user and eavesdropper channel qualities $0\leq(\gamma^e, \bar{\gamma})<1$. Suppose that the encoding design variable $\mu$ in \eqref{eq:muchoice} is chosen in the range
\begin{equation*}
    \mu^{OP} < \mu < 1,
\end{equation*}
where $ \mu^{OP}$ is given in \eqref{eq:mudesignforopenloop}.
Then $\lim_{k\to\infty}\mathbb{E}[P_k] < P^{OP}$ and $\lim_{k\to\infty}\mathbb{E}[P_k^e] > P^{OP}$.
\end{thm}
\begin{pf}
The proof is included in the Appendix.
\end{pf}
In comparison with the state secrecy codes developed for stable systems \citep{tsiamis2018state}, the current method allows one to drive the eavesdropper's estimation error variance above open loop prediction through transmitting noise. Thus, interceptions of packets are not informative to the eavesdropper's estimation process. Moreover, the eavesdropper intercepting packets is actively harming its own estimation performance. Since sending noise harms the legitimate user's performance as well, we choose $\mu < 1$ so the performance of the legitimate user remains upper-bounded by open loop performance. This ensures the interceptions of packets remain informative to the legitimate user.

\section{Examples} \label{sec:examples}
In this section, we illustrate the effectiveness of our encoding scheme. We consider a first order system \eqref{eq:lti} with parameters $a = 0.6$, $q =0.01$, $r = 0.01$ and the probability of channel dropout for the eavesdropper $\gamma^e = 0.3$ and legitimate user $\bar{\gamma} = 0.3$. This gives $P_n = 0.0256$, $\bar{P} = 0.0054$ and $P^{OP} = 0.0156$.
\begin{figure}
    \centering
    \includegraphics[width=8.4cm]{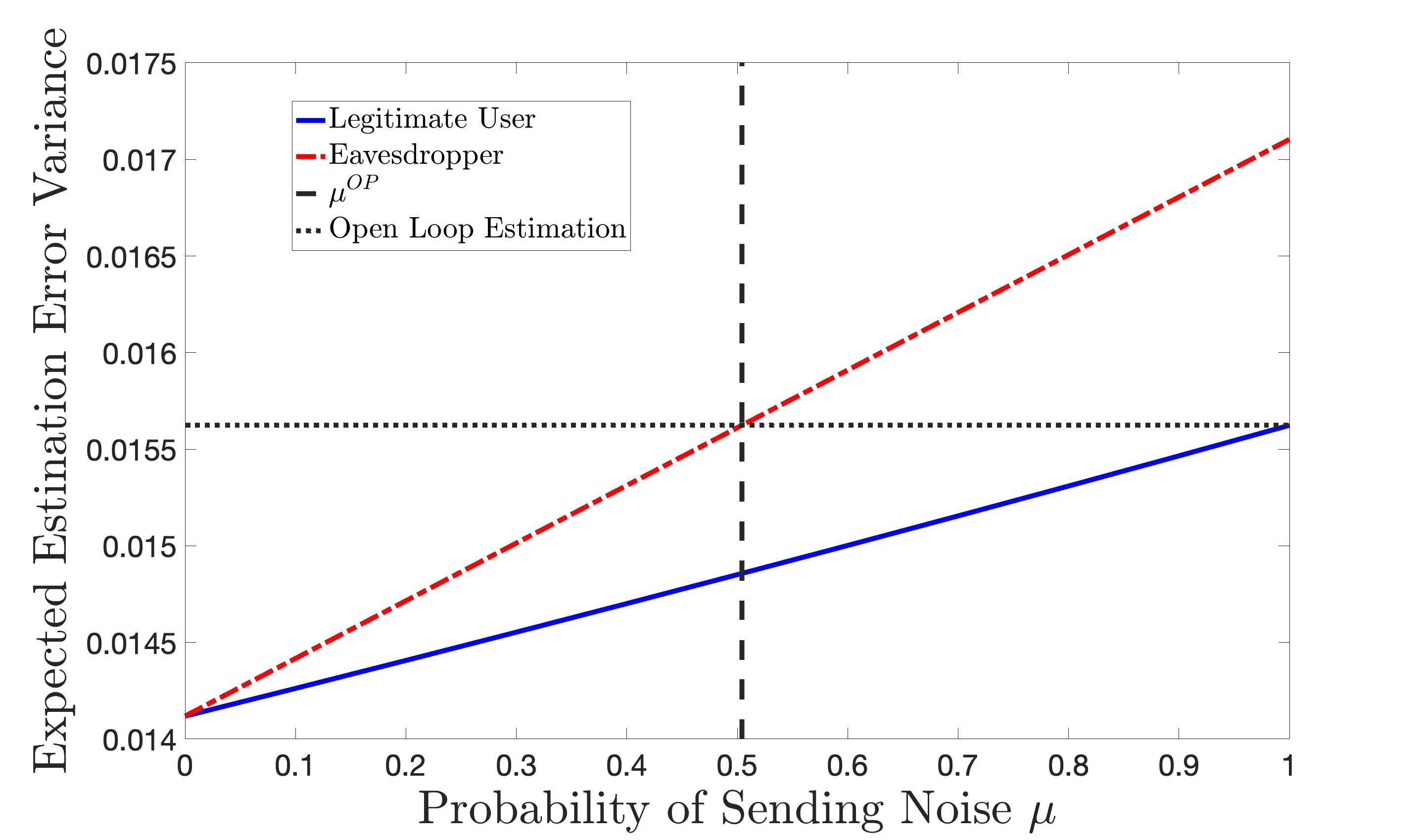}
    \caption{Expected long-run estimation error variance for the eavesdropper and legitimate user. The eavesdropper's expected estimation error variance can be greater than open loop. The legitimate user's estimation error variance remains bounded by open loop prediction. A value of $\mu$ is chosen that balances the trade-off between data confidentiality and performance degradation for the legitimate user.}
    \label{fig:expectedcov}
\end{figure}
Fig.\ \ref{fig:expectedcov} gives the numerical expectation of the long-run estimation error variance for the legitimate user \eqref{eq:esterrorlegitimatefull} and eavesdropper \eqref{eq:esterroreavesdropperfull} for a probability of the sensor sending noise $\mu = (0, 1)$.
For all valid choices of $\mu$,
the expected long-run estimation error variance of the eavesdropper is greater than the legitimate user's. 
We compute the minimum choice of $\mu$ that will drive the eavesdropper's expected estimation error larger than open-loop performance from \eqref{eq:mudesignforopenloop}, with $\mu^{OP} = 0.504$, denoted by a vertical dotted line. We denote the open loop prediction value, $P^{OP}$ with a horizontal dotted black line.
Following Theorem~\ref{thm:secrecy}, the encoding design variable $\mu$ in the range $0.504 < \mu < 1 $ drives the eavesdropper's expected estimation error above open loop prediction, whilst keeping the legitimate user's expected estimation error upper-bounded by open loop prediction. 

\section{Conclusion} \label{sec:conclusion}
This paper has developed an encoding scheme that ensures confidentiality of the state of a scalar system, protecting data confidentiality, in a network environment without packet receipt acknowledgments.
Under our encoding design, the expected performance of the eavesdropper is larger than the legitimate user. For certain design choices, the eavesdropper's expected performance is worse than the open loop estimation performance, whilst the legitimate user's expected performance is better than the open loop estimation performance. An open problem is the inclusion of a scheme to detect the possible presence of an eavesdropper to activate the proposed encoding scheme. Additionally, it remains an open problem to extend this encoding scheme beyond first-order systems.
\bibliography{ifacconf}

\begin{thebibliography}{12}
\providecommand{\natexlab}[1]{#1}
\providecommand{\url}[1]{\texttt{#1}}
\providecommand{\urlprefix}{URL }
\expandafter\ifx\csname urlstyle\endcsname\relax
  \providecommand{\doi}[1]{doi:\discretionary{}{}{}#1}\else
  \providecommand{\doi}{doi:\discretionary{}{}{}\begingroup
  \urlstyle{rm}\Url}\fi

\bibitem[{Anderson and Moore(1979)}]{anderson2012optimal}
Anderson, B.D. and Moore, J.B. (1979).
\newblock \emph{Optimal filtering}.
\newblock Englewood Cliffs, N.J., USA: Prentice-Hall.

\bibitem[{Br{\'e}maud(1999)}]{bremaud2013markov}
Br{\'e}maud, P. (1999).
\newblock \emph{Markov chains}.
\newblock New York, NY, USA: Springer-Verlag.

\bibitem[{Cardenas et~al.(2008)Cardenas, Amin, and Sastry}]{cardenas2008secure}
Cardenas, A.A., Amin, S., and Sastry, S. (2008).
\newblock Secure control: Towards survivable cyber-physical systems.
\newblock In \emph{28th International Conference on Distributed Computing
  Systems Workshops}, 495--500. Berkeley, California.

\bibitem[{Ding et~al.(2017)Ding, Li, Quevedo, Dey, and Shi}]{ding2017multi}
Ding, K., Li, Y., Quevedo, D.E., Dey, S., and Shi, L. (2017).
\newblock A multi-channel transmission schedule for remote state estimation
  under {DoS} attacks.
\newblock \emph{Automatica}, 78, 194--201.

\bibitem[{Ding et~al.(2021)Ding, Ren, Leong, Quevedo, and Shi}]{ding2020remote}
Ding, K., Ren, X., Leong, A.S., Quevedo, D.E., and Shi, L. (2021).
\newblock Remote state estimation in the presence of an active eavesdropper.
\newblock \emph{IEEE Transactions on Automatic Control}, 66(1), 229--244.

\bibitem[{Humayed et~al.(2017)Humayed, Lin, Li, and Luo}]{humayed2017cyber}
Humayed, A., Lin, J., Li, F., and Luo, B. (2017).
\newblock Cyber-physical systems security—a survey.
\newblock \emph{IEEE Internet of Things Journal}, 4(6), 1802--1831.

\bibitem[{Ishii and Zhu(2022)}]{ishii2022security}
Ishii, H. and Zhu, Q. (eds.) (2022).
\newblock \emph{Security and Resilience of Control Systems}.
\newblock Springer.

\bibitem[{Katz and Lindell(2020)}]{katz2020introduction}
Katz, J. and Lindell, Y. (2020).
\newblock \emph{Introduction to modern cryptography}.
\newblock CRC press.

\bibitem[{Klingler and Dressler(2019)}]{klingler2019jamming}
Klingler, F. and Dressler, F. (2019).
\newblock Jamming {WLAN} data frames and acknowledgments using commodity
  hardware.
\newblock In \emph{IEEE Conference on Computer Communications Workshops},
  1015--1016. Paderborn, Germany.

\bibitem[{Lee et~al.(2010)Lee, Kapitanova, and Son}]{lee2010price}
Lee, J., Kapitanova, K., and Son, S.H. (2010).
\newblock The price of security in wireless sensor networks.
\newblock \emph{Computer Networks}, 54(17), 2967--2978.

\bibitem[{Naha et~al.(2022)Naha, Teixeira, Ahl{\'e}n, and
  Dey}]{naha2022quickest}
Naha, A., Teixeira, A., Ahl{\'e}n, A., and Dey, S. (2022).
\newblock Quickest detection of deception attacks on cyber-physical systems
  with a parsimonious watermarking policy.
\newblock \emph{arXiv:2201.09389}.

\bibitem[{Tsiamis et~al.(2018)Tsiamis, Gatsis, and Pappas}]{tsiamis2018state}
Tsiamis, A., Gatsis, K., and Pappas, G.J. (2018).
\newblock State-secrecy codes for stable systems.
\newblock In \emph{2018 American Control Conference (ACC)}, 171--177.
  Milwaukee, USA.

\end{thebibliography}
\appendix
\section{Proof of Results}
\subsection{Proof to Lemma \ref{lemma:Estimation error variance}}
 Consider the equation for the eavesdropper's state estimation process in \eqref{eq:eveasdropperestimation}. To prove \eqref{eq:esteavesdropper}, we firstly consider the scenario that the eavesdropper successfully receives the state estimate from the sensor with $(\lambda_k^e,u_k) = (1,1)$, the eavesdropper's estimate is $\hat{x}_k^e=\hat{x}_k^s$ and the estimation error variance is that of the sensor's \eqref{eq:pbar} with $P_k^e = \bar{P}$. Secondly, in the scenario that a packet dropout occurs with  $\lambda_k^e = 0$, the eavesdropper predicts $\hat{x}_k^e$ with that of its previous estimate, with $\hat{x}_k^e=a\hat{x}_{k-1}^e$. The corresponding estimation error variance is $P_{k}^e = a^2P_{k-1}^e + q$. Thirdly, in the scenario that the eavesdropper successfully receives a packet of noise $(\lambda_k^e,u_k) = (1,0)$, the eavesdropper directly uses the packet as the state estimate $\hat{x}_k^e=n_k$. The estimation error variance is
\begin{equation} \label{eq:covuppereavesdropperexpanded}
P_n = \mathbb{E}[(x_k-n_k)^2| \mathcal{I}_k^e] =  \mathbb{E}[x_k^2] + 2\mathbb{E}[x_kn_k] + \mathbb{E}[n_k^2] .
\end{equation}
These cases show \eqref{eq:esteavesdropper}. By definition the transmitted noise is designed such that $\mathbb{E}[n_k^2] = q$, uncorrelated with $x_k$ such that $\mathbb{E}[x_kn_k] = 0$.
As per \eqref{eq:openloop}, the open loop estimation error variance is $\mathbb{E}[x_k^2] = P^{OP}$. Then from \eqref{eq:covuppereavesdropperexpanded}
\begin{equation*}
    P_n = \mathbb{E}[x_k^2] + 2\mathbb{E}[x_kn_k] + \mathbb{E}[n_k^2] = P^{OP} + 0 + q,
\end{equation*}
which shows \eqref{eq:covuppereavesdropper}.

\subsection{Proof to Lemma \ref{lemma:estimation error legitimate}}
The mathematical expectation of the legitimate user's estimation error variance at time $k$ can be found by taking the sum of all possible channel outcomes multiplied by the corresponding probability of that channel realisation occurring.
Let $S_k$ be the state of a Markov chain taking values in the countably infinite set $S_k \in \{0,1,\dots\}$.
We define $S_k = 0$ as the state when the sensor's state estimate is received, and $S_k = j$ for $j > 0$ as the $j$th dropout.
By application of \eqref{eq:legitchannel} and \eqref{eq:muchoice} 
the transition probability $p_{ij} \triangleq \mathbb{P}[S_{k+1} = i |S_{k} = j]$ is characterized by
\begin{equation*}
    \mathbb{P}[S_{k+1} = i |S_{k} = j] = 
    \begin{cases} 
    \bar{\gamma} + (1-\bar{\gamma})\mu, & \text{ if $i =  0$}, \\
    (1-\bar{\gamma})(1-\mu), & \text{ if $i = j+1$}, \\
    0,  & \text{ otherwise}.
    \end{cases}
\end{equation*}
We define the transition matrix of the Markov chain as $\mathbf{P} = [p_{ij}]$.
Since all states communicate, the Markov chain is irreducible, aperiodic, and recurrent and has a stationary distribution $\pi$ \citep{bremaud2013markov}.
By solving the relation $\pi = \pi \mathbf{P}$ with $\sum_{j=0}^{\infty}\pi_j = 1$, the stationary distribution can be found as 
\begin{equation*} \label{eq:stationarylegitimate}
\begin{aligned}
\pi_j &= (\bar{\gamma} + (1-\bar{\gamma})\mu)^j(1-\bar{\gamma})(1-\mu).
\end{aligned}
\end{equation*}
Following \eqref{eq:estlegitimate}, the conditional expectation of the estimation error variance $P_k$ at Markov chain state $S_k = j$ is 
\begin{equation*}
    \mathbb{E}[P_k|S_k = j] =  (a^2)^{j}\bar{P} + \sum_{i=0}^{j - 1} (a^2)^{i}q . 
\end{equation*}
Taking the limit as $k \rightarrow \infty$, then the expected estimation error variance is
\begin{equation*}
\begin{split}
    &\lim_{k\to\infty}\mathbb{E}[P_k] =  \lim_{k\to\infty}\sum_{j=0}^{k} \mathbb{E}[P_k|S_k = j]\pi_j = \\
    &(1-\bar{\gamma})(1-\mu) \sum_{j=0}^{\infty}(\bar{\gamma} + (1-\bar{\gamma})\mu)^j \left((a^2)^{j}\bar{P} + \sum_{i=0}^{j - 1} (a^2)^{i}q \right) .
\end{split}
\end{equation*}
As $\bar{\gamma} < 1$ and $\mu < 1$, and $a$ is stable and scalar, this expression reduces into \eqref{eq:esterrorlegitimatefull}.
The sum of an infinite number of terms that have a constant ratio between successive terms reduces into a geometric series, where $(1-\bar{\gamma})(1-\mu) \sum_{j=0}^{\infty}(\bar{\gamma} + (1-\bar{\gamma})\mu)^j (a^2)^{j}\bar{P}$ reduces to $\frac{\bar{P}(1-\bar{\gamma})(1-\mu) }{1 - a^2(\bar{\gamma} + (1-\bar{\gamma})\mu)}$ and by expanding the summation $(1-\bar{\gamma})(1-\mu)\sum_{j=0}^{\infty}(\bar{\gamma} + (1-\bar{\gamma})\mu)^j\sum_{i=0}^{j-1}(a^2)^{i}q = (\bar{\gamma} + (1-\bar{\gamma})\mu)\sum_{j=0}^{\infty}(\bar{\gamma} + (1-\bar{\gamma}))^j(a^2)^jq$ which reduces to $\frac{q(\bar{\gamma} + (1-\bar{\gamma})\mu)}{1 - a^2(\bar{\gamma} + (1-\bar{\gamma})\mu)}$.
If $\bar{\gamma} = 1$ then no packets are received, or if $\mu = 1$ when only noise is sent then the legitimate user receives no packets containing the state, as such the Markov chain does return to state $S_k=0$ and the estimation error variance satisfies
\begin{equation} \label{eq:open loop legitimate}
    \lim_{k\to\infty}\mathbb{E}[P_k] = P^{OP}, \text{ if $\bar{\gamma} = 1$ or $\mu = 1$}.
\end{equation}

\subsection{Proof to Lemma \ref{lemma:estimation error eavesdropper}}
\begin{figure}
    \centering
    \includegraphics[width=8.4cm]{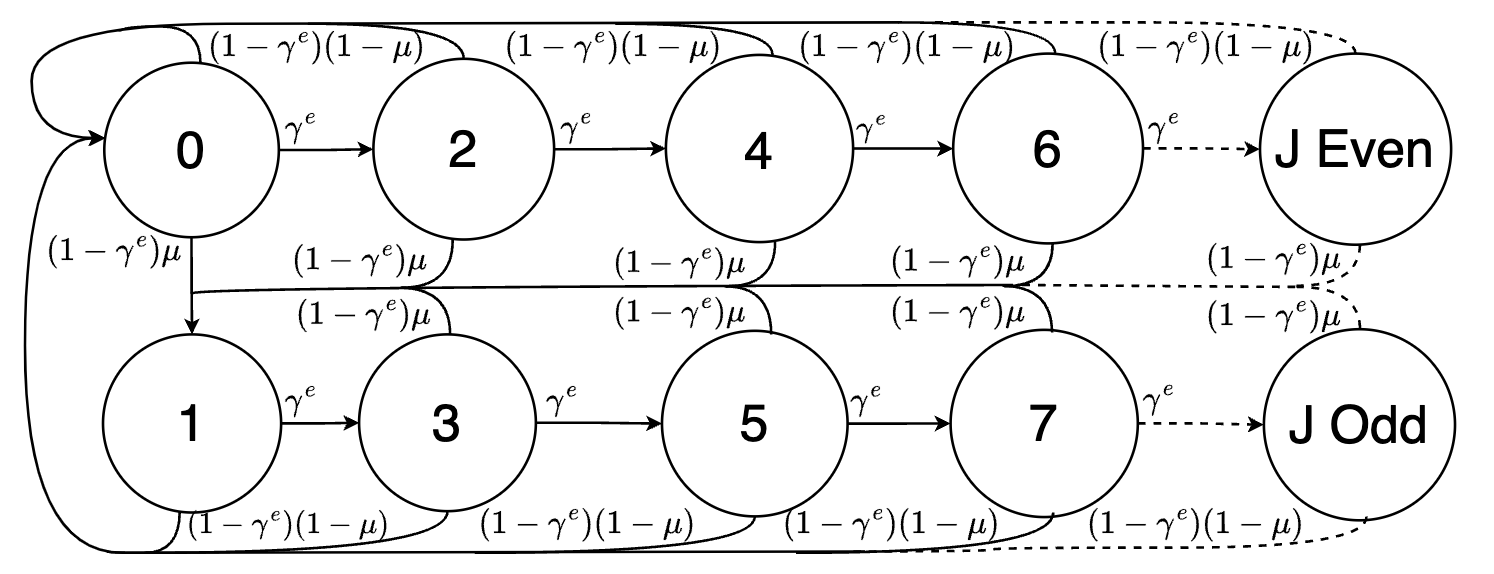}
    \caption{Markov chain model for eavesdropper.} 
    \label{fig:eavesdropperchain}
\end{figure}

The mathematical expectation of the eavesdropper's estimation error variance at time $k$ can be found by taking the sum of all possible channel outcomes and transmission sequences multiplied by the corresponding probability of that realisation occurring.
Let $S_k^e$ be the state of a countably infinite Markov chain taking values in the  set $S_k^e \in \{0,1,\dots\}$. We define two key events, $S_{k}^e=0$ as the state when the sensor's state estimate is received, and $S_{k}^e=1$ as the state when noise is received by the eavesdropper.
These two events can be followed by an infinite number of packet dropouts.
Let $S_k^e=j$ for $j \geq 2$, be the dropouts after $S_\ell^e=0$ when $j$ is even, and $S_\ell^e=1$ when $j$ is odd, for some $\ell < k$.
The states $S_{k}^e=0$ and $S_{k}^e=1$ can be reached from any other state.
By application of \eqref{eq:eavesdropperchannel} and \eqref{eq:muchoice} 
the transition probability $p_{ij}^e \triangleq \mathbb{P}[S_{k+1}^e = i |S_{k}^e = j]$ is characterized by
\begin{equation*}
     \mathbb{P}[S_{k+1}^e = i |S_{k}^e = j] = 
    \begin{cases} 
    (1-\gamma^e)(1-\mu), & \text{ if $i = 0$,} \\
    (1-\gamma^e)\mu, & \text{ if $i =  1$,} \\
    \gamma^e,  & \text{ if $i = j + 2$}, \\
    0, & \text{ otherwise .}
    \end{cases} 
\end{equation*}
The Markov chain and its transitions are depicted in Fig. \ref{fig:eavesdropperchain}.
We define the transition matrix of the Markov chain as $\mathbf{P}^e = [p_{ij}^e]$.
Since all states communicate, the Markov chain is irreducible, aperiodic, and recurrent and has a stationary distribution $\pi^e$ \citep{bremaud2013markov}.
By solving the relation $\pi^e = \pi^e\mathbf{P}^e$ with $\sum_{j=0}^{\infty}\pi_j^e = 1$, the stationary distribution for all states are as follows 
\begin{align*}
    \pi_0^e &= (1-\gamma^e)(1-\mu), ~ \pi_1^e = (1-\gamma^e)\mu, \mbox{ and }
    \pi_{j+2}^e = (\gamma^e)\pi_j^e .
\end{align*}
Following \eqref{eq:esteavesdropper}, the conditional expectation of the estimation error variance $P_k^e$ at Markov chain state $S_k^e = j$ is 
\begin{align*}
    &\mathcal{P}_j^e = \mathbb{E}[P_k^e|S_k^e = j] \\
    & = \begin{cases}
        (a^2)^{\frac{1}{2}j}\Bar{P} + \sum_{i=0}^{\frac{1}{2}j - 1} (a^2)^{i}q,&\text{ if $j$ even},\\ 
        (a^2)^{\frac{1}{2}(j-1)}P_n + \sum_{i=0}^{\frac{1}{2}(j-1) - 1} (a^2)^{i}q,&\text{ if $j$ odd} . 
    \end{cases}
\end{align*}
Taking the limit as $k\rightarrow\infty$, then the expected estimation error variance is
\begin{equation*}
\begin{split}
    &\lim_{k\to\infty}\mathbb{E}[P_{k}^e] = \lim_{k\to\infty}\sum_{j=0}^{k} \mathbb{E}[P_k^e|S_k^e = j]\pi_j^e \\
    &= \sum_{j=0}^{\infty} (\gamma^e)^j \pi_0^e \mathcal{P}_{2j}^e + \sum_{j=0}^{\infty}(\gamma^e)^j \pi_1^e \mathcal{P}_{2j + 1}^e \\
    &= (1-\gamma^e)(1-\mu) \sum_{j=0}^{\infty}(\gamma^e)^j\left((a^2)^{j}\bar{P} + \sum_{i=0}^{j - 1} (a^2)^{i}q\right)  \\  &\quad+ (1-\gamma^e)\mu\sum_{j=0}^{\infty}(\gamma^e)^j\left((a^2)^{j}P_n + \sum_{i=0}^{j - 1} (a^2)^{i}q\right) .
\end{split}
\end{equation*}
As $\gamma^e < 1$ and $a$ is stable and scalar, this expression reduces into \eqref{eq:esterroreavesdropperfull},  as the sum of an infinite number of terms that have a constant ratio between successive terms reduces into a geometric series, where $(1-\gamma^e)(1-\mu)\sum_{j=0}^{\infty}(\gamma^e)^j(a^2)^{j}\bar{P}$ reduces to $\frac{\bar{P}(1-\gamma^e)(1-\mu) }{1 - a^2\gamma^e}$ and expanding the summation $(1-\gamma^e)\mu\sum_{j=0}^{\infty}(\gamma^e)^j \sum_{i=0}^{j - 1} (a^2)^{i}q = \mu\gamma^e\sum_{j=0}^{\infty}(\gamma^e)^j(a^2)^{j}q$ which reduces to $\frac{q\mu\gamma^e}{1 - a^2\gamma^e}$. We also have $(1-\gamma^e)(1-\mu)\sum_{j=0}^{\infty}(\gamma^e)^j  \sum_{i=0}^{j-1}(a^2)^{i}q  = \gamma^e(1-\mu)\sum_{j=0}^{\infty}(\gamma^e)^j(a^2)^jq$ by expanding the summation, which reduces to  $\frac{q\gamma^e(1-\mu)}{1 - a^2\gamma^e}$, with  $(1-\gamma^e)\mu\sum_{j=0}^{\infty}(\gamma^e)^j(a^2)^{j}P_n$ reducing to $\frac{ P_n(1-\gamma^e)\mu}{1 - a^2\gamma^e}$. If $\gamma^e = 1$, then there are only dropouts and the Markov chain cannot return to state $S_k^e = 0$ or $S_k^e = 1$ for any $k > 0$ and the estimation error variance satisfies 
\begin{equation*}
    \lim_{k\to\infty}\mathbb{E}[P_{k}^e] = P^{OP} \text{ if $\gamma^e = 1$}.
\end{equation*}

\subsection{Proof to Lemma \ref{lemma:mu}}
Given the upper bound for the eavesdropper estimation error in \eqref{eq:covuppereavesdropper}, we can derive the conditions that achieve this through re-arranging the eavesdropper's expected long-run estimation error in \eqref{eq:esterroreavesdropperfull} in terms of $\mu $.
We desire that $\lim_{k\to\infty}\mathbb{E}[P_k^e] = P^{OP}$, with the probability of sending noise that achieves this denoted as $\mu^{OP}$, given that
\begin{equation*}
    \mu^{OP} = \frac{P^{OP}(a^2\gamma^e -1) -\gamma^e\bar{P} + \gamma^eq + \bar{P}}{(\gamma^e -1)(P_n - \bar{P})} .
\end{equation*}
\subsection{Proof to Theorem \ref{thm:secrecy}}
To drive the eavesdropper's expected estimation error variance to the open loop variance, we would choose $\mu = \mu^{OP}$.
The legitimate user's expected estimation error variance has an upper bound of $P^{OP}$ when $\mu = 1$ by \eqref{eq:open loop legitimate}.
A choice of $\mu < 1$ ensures that $\lim_{k\to\infty}\mathbb{E}[P_k] < P^{OP}$, and $\mu > \mu^{OP}$ ensures that $\lim_{k\to\infty}\mathbb{E}[P_k^e] > P^{OP}$,
giving the range of $\mu$ of $\mu^{OP} < \mu < 1$.

\end{document}